# COMS-Integrated Atomic Vapor Cells with Ultra-long Optical Access for Highly Sensitive and Scalable Quantum Sensors

Yintao Ma[a,d], Yao Chen[a,d*], Mingzhi Yu[a,d], Yanbin Wang[a,c], Ju Guo[a,c], Ping Yang[a,c], Qijing Lin[a,b,d], Yang Lv[e], Libo Zhao[a,b,d*]

[a] State Key Laboratory for Manufacturing Systems Engineering, State Industry-Education Integration Center for Medical Innovations, International Joint Laboratory for Micro/Nano Manufacturing and Measurement Technologies, Shaanxi Innovation Center for Special Sensing and Testing Technology in Extreme Environments, Shaanxi Provincial University Engineering Research Center for Micro/Nano Acoustic Devices and Intelligent Systems, Xi'an Jiaotong University, Xi'an 710049, China

[b] Shandong Laboratory of Advanced Materials and Green Manufacturing at Yantai, Yantai 264000, China

[c] School of Mechanical Engineering, Xi'an Jiaotong University, Xi'an 710049, China

[d] School of Instrument Science and Technology, Xi'an Jiaotong University, Xi'an 710049, China

[e] Zhuzhou CRRC Times Electric Co., Ltd, Zhuzhou 412000, China

*corresponding authors (E-mail: chenyao@xjtu.edu.cn and libozhao@xjtu.edu.cn)

The most appealing features of chip-scale quantum sensors are their capability to maintain extreme sensitivity while enabling large-scale batch manufacturing. This necessitates high-level integration and wafer-level fabrication of atomic vapor cells. In this paper, we describe a micromachining paradigm for wafer-level atomic vapor cells functionalized by CMOS-compatible non-magnetic heaters and temperature sensors and demonstrate several innovative applications. Leveraging standard micro-nanofabrication technology, the integrated vapor cells achieved an ultra-long optical access of 5 mm, nearly four time that of previously microfabricated vapor cells. The feasibility of the integrated atomic vapor cells fabrication process was verified by a consecutive 30-day aging test in a harsh environment (operating temperature of 200 °C and vacuum of approximately 1 Pa). Benefiting from the ultra-long optical path, we observed several typical quantum effects, including the saturation absorption and spin fluctuations, a regime previously inaccessible with conventional micromachined vapor cells. Finally, a zero-field quantum magnetometry with an ultra-high magnetic sensitivity of 12 fT/Hz$^{1/2}$ was also demonstrated. Our achievements broaden the potential applications of microfabricated atomic vapor cells and pave the way for scalable manufacturing of ultrasensitive, chip-scale quantum sensors.

**1. Introduction**

Chip-scale quantum sensors, empowered by micro-electro-mechanical system (MEMS) technologies, have versatile application prospects in the fields of navigation positioning and timing [1, 2], cutting-edge physics experiments [3-5], bio-magnetic field imaging [6-8], and beyond. Among all quantum sensors, a category of quantum sensing systems based on alkali-metal vapor atoms or their derivatives, such as atomic magnetometers, Rydberg atoms, and atomic spin gyroscopes, have recently gained overwhelming interest owing to their advantages of ultra-high measurement sensitivity and susceptibility to polarization by commercially available lasers. However, the highly strong reactivity of alkali metal atoms necessitates that they must be confined inside an enclosed and transparent container, known as the atomic vapor cells (AVCs).

The AVCs are commonly manufactured utilizing the glass-blowing method [9, 10], which would undoubtedly result in issues such as low efficiency, poor consistency, and high cost. To address these existing shortcomings, micro-machined AVCs were subsequently developed as an alternative and promising technology, enabled by the standard MEMS fabrication process, in particular silicon micro-machining and anodic bonding procedures [11-13]. One has to acknowledge that the successful implementation of MEMS-based AVCs has been a significant catalyst for the leapfrog development of microfabricated quantum sensors. Although the advancements in micromachining AVCs have been ongoing for more than twenty years, to date, the vast majority of previous demonstrations of microfabricated thermal vapor atom-based quantum sensors rely on MEMS vapor cells with an effective optical access of approximately 1.5 mm [14-16] or separate stand-alone components (flexible heating films and commercially available temperature sensors) [17-20]. This limited optical access and lack of all-in-one functionality decreases the sensitivity of quantum sensors and obstacles the on-chip integration.

Indeed, the AVCs endowed with longer optical path and integrated with more functionalities at the wafer-level are more favorable for chip-scale quantum sensors owing to more effective vapor atoms and more compact package, which provides enhanced sensitivity as well as the ability to engineer on-chip. With this general idea in mind, the MEMS-based AVCs are developing in two directions: increased optical length and versatility. Three approaches are typically used to increase the effective optical length of AVCs. The first approach is to drill through-holes on thick silicon/glass wafers using conventional machining techniques [21-23], such as mechanical drilling and water jet processes, which, in addition to contaminating the wafer surface, would lead to chipping of the holes, thus imposing a detrimental effect on subsequent anodic bonding and hermetic sealing. As a derivative, sidewalls with integrated reflectors are fabricated on the silicon body by wet etching [24], allowing the laser beam to be reflected twice, thereby prolonging the optical length. The difficulties in controlling wet etching and the complexity of the structure make this method unsuitable for wafer-level precision fabrication. Besides that, a combination of multi-pass configuration or optical resonant cavity with AVCs would also notably increase the effective optical length [25-27]. However, the two major constraints are the lack of miniaturization and the highly environmentally sensitive nature of these devices. Although ultra-long optical path MEMS AVCs with all-in-one functionalized components, such as non-magnetic heaters and temperature sensors, are valuable and desirable, no related reports have been presented yet.

In this study, we developed CMOS-integrated, wafer-level AVCs with ultra-long optical path by means of the standard MEMS technologies. In this process, double-sided dry etching with hard masks and low-temperature anodic bonding were utilized, and the functionalized components were then assembled at the wafer level to form all-in-one vapor cells. Using the microfabricated vapor cells, we demonstrated several novel quantum effects benefiting from the physical features of the ultra-thick silicon core, including the saturation absorption spectrum and spin noise spectrum, in addition to configuring an ultra-sensitive zero-field atomic magnetometer.

**2 Methods**

2.1. Microfabrication process

Fig. 1 shows the microfabrication process of the CMOS-integrated functionalized MEMS AVCs. The process begins with a p-type <100>-oriented silicon wafer with a thickness of 5 mm and two borosilicate glass wafers (BF33) with a thickness of 500 μm.

The detailed procedure is depicted as follows:

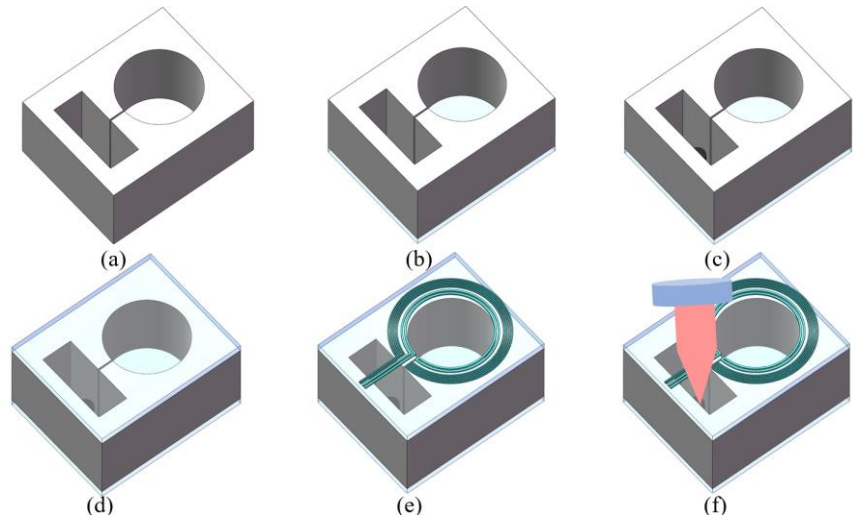

Fig. 1. Fabrication process of the CMOS-integrated functionalized MEMS AVCs. (a) ICP etching of silicon through-holes; (b) First anodic bonding to make preformed cavities; (c) Injection of alkali metal atoms; (e) Second anodic bonding to seal the AVCs; (e) Deposition of Pt metal for heating and temperature sensing; (f) Activation of the dispenser.

(a) Etching of silicon through-holes

  A double-sided polished <100>-oriented silicon wafer with a thickness of 5 mm was used as the substrate, and the 600 nm Al layer was deposited as the mask layer on the top and bottom surfaces of the silicon wafer by magnetron sputtering. The pattern of the AVCs was then transferred from the photomask to the silicon substrate by photolithography, following which the dry etching process removed excess silicon material using the inductively coupled plasma (ICP) technique, which involved both physical bombardment and chemical corrosion, until the silicon through-holes were achieved.

(b) Preformed cavities created by first anodic bonding

  The silicon wafer with etched through-holes was subsequently anodically bonded to a 500 μm-thick borosilicate glass BF33 wafer of the same size by heating the wafers to a temperature of 350 °C and applying a voltage of 1000 V across the silicon and glass wafers. Together, the combination of heat and a high electric field enabled the formation of Si-O chemical bonds at the interface, resulting in hermetically sealed bonds between the silicon and glass wafers. Hereafter, we refer to the obtained silicon-glass stack as "the preformed cavities".

(c-d) Injection of Rb atoms and sealing of the AVCs

  The Rb dispenser pills purchased from SAES Getters consisting of Rb chromate and Zr–Al alloy mixture, were then distributed one by one into the preformed cavities. The dispenser was chemically stable up to 500 °C and could be activated at higher temperatures to generate Rb vapor atoms. Subsequently, a second BF33 glass wafer same as used for the first anodic bonding was bonded to the "preformed cavities" using plasma-activated low-temperature anodic bonding to enclose the AVCs in a gas-controlled atmosphere. The wafer surfaces were treated with $O_2$ and Ar plasma activation to perform low-temperature bonding. The mixed plasma was used for the activation modification of the wafer surfaces, resulting in a temperature as low as 150 °C for the second anodic bonding, which was compatible with the operating temperature of octadecyltrichlorosilane (OTS). Low-temperature anodic bonding was selected to seal the AVCs for the following reasons: Firstly, localized increases in temperature during the bonding process may result in a loss of buffer gas owing to the gas molecule diffusion, which leads to a discrepancy between the actual filled gas pressure and the final gas pressure inside the cells. Secondly, a hermetic sealing temperature below 170°C provides a possible solution to the problem that the conventional anodic bonding temperature is unavailable to OTS anti-relaxation coatings. Finally, the glass-silicon-glass sandwich-structured AVCs have optically transparent windows for optical propagation on the top and bottom surfaces.

(e) Functionalized components in optically transparent windows

A temperature-controlled chip (including heaters and temperature sensors) is composed of two identical layers of serpentine-shaped resistance separated by an isolated layer in the middle, and electrically connected through a through-hole between the two layers. Each layer has two sets of wires serving as a temperature sensor and heating resistor, respectively, which together enable a closed-loop temperature control system. Note that the heater resistor structure is centered on the optical chamber, to provide a higher temperature in the optical chamber than that in the reservoir chamber. This will facilitate the prevention of the formation of condensed alkali-metal atoms droplets in the optical chamber, which would impede the optical access as well as resulting in extra thermal magnetic noise.

The fabrication of the heating chip was divided into three steps: Firstly, Pt resistors of the first layer were accessed sequentially via photolithography, magnetron sputtering, and lift-off. A 400 nm silicon-oxide insulating layer was then deposited on the first metal layer, with an electrical connection established via ICP etching. Finally, a second layer of resistors was fabricated using the same process as the first layer, thereby forming complete all-in-one integrated AVCs.

(f) Activation of the dispenser

The dispenser was activated at a high temperature by focusing a 3 W laser beam with a wavelength of 1550 nm on the dispenser, which absorbed heat and generated a localized high temperature. The activation generated silver white Rb atoms that could be visualized after the dispenser was exposed to the beam for approximately 5 s.

Eventually, 10 completed microfabricated heating chips were arbitrarily selected, and the resistance measured using the current ($I$)–voltage ($V$) characteristic curve of the semiconductor device parameter analyzer showed an average value of heating resistance of 505 Ω and temperature sensor resistance of approximately 10.5 kΩ, as shown in Fig. 2, which closely matched the designed value.

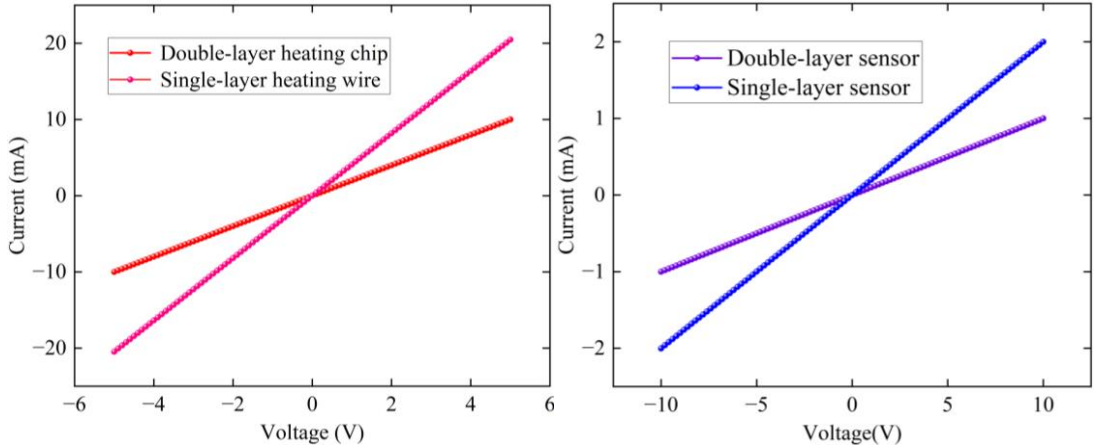

Fig. 2. *I-V* characteristic curve of the temperature sensors and heating resistors.

2.2. Experimental setup

We utilized the experimental setup shown in Fig. 3 to implement saturated absorption spectroscopy (SAS), which was applicable to laser frequency stabilization, spin noise spectroscopy (SNS); a non-destructive spin fluctuations measurement technique; and a single-beam zero-field optical atomic magnetometry, which is a typical example of atomic quantum sensor. The apparatus consists of three key components: (1) the CMOS-integrated functionalized AVCs, including Rb atoms, a heater, and a temperature controller, as the core component of the quantum sensor; (2) the laser used to optically pump and probe the atomic spin; and (3) the magnetic shielding and coils used for magnetic compensation, magnetic modulation, and tuning Rb atoms Larmor precession frequency.

An integrated MEMS cell with an effective optical length of 4 mm was at the heart of the experimental setup. The cell contained natural Rb atoms, as well as pressure-tunable $N_2$ acting both as the quenching gas to prevent radiation trapping effect, and as the buffer gas to slow the diffusion motion to the cell walls. To ensure the availability of a large number of vapor atoms from the saturated vapor pressure, the cell was AC electrically heated with the use of heaters integrated on the transparent surface of the cell to suppress the undesired magnetic noise. In addition to the heaters, the temperature of the cell was real-time monitored using the assembled temperature sensors.

The CMOS-integrated cell was encapsulated in a field-controlled environment comprising four layers of magnetic shielding and three-dimensional (3D) magnetic field coils for the propose of generating the intended magnetic field. A laser beam with a central frequency in the D1 line of Rb atoms was delivered to the optical chamber of the cell via a single-mode polarization maintaining fiber for interaction with the atomic vapor. For the SAS experiment, the beam transmitted through the cell was reflected by a planar mirror and passed back again before hitting the photodetector. However, the transmitted beam after cell was, in the case of SNS and single-beam atomic magnetometry, analyzed by a balanced polarimeter, which was composed of a half waveplate, a polarizing beam splitter and two photodetectors.

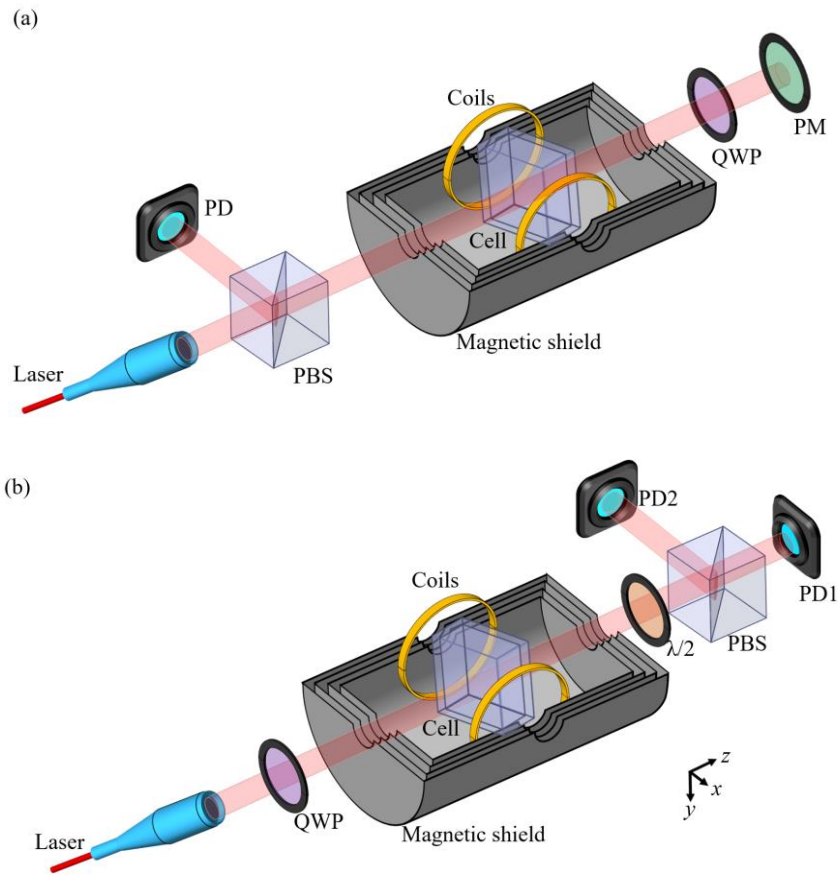

Fig. 3. Simplified experimental setup. (a) Saturated absorption spectroscopy (SAS). The SAS experiment was arranged as a double-pass layout consisting of a quarter wave plate (QWP), a reflecting plane mirror (PM), and a photodetector (PD) after twice passing through the functionalized atomic vapor cell. (b) Spin noise spectroscopy (SNS) and single-beam atomic magnetometer. The two experiments adopted a single-pass layout featuring a QWP in front of the atomic interactions and a balanced polarimeter comprising a half waveplate ($\lambda/2$), a polarizing beam splitter (PBS), and magnified differential photodetectors (PD1 and PD2).

### 3. Experimental Results
3.1. Heating chip testing

We first conducted comprehensive performance tests on the integrated heating chip, including the heating characteristics, temperature stability, and the residual magnetic flux density arising from the heating process. The testing results are drawn in Fig.4. The temperature was precisely regulated through proportional–integral–derivative (PID)-controlled voltage adjustment of the heating resistor enabled by a LabVIEW control program. Fig. 4(a) illustrates the heat-up characteristic curve and temperature stability at a setpoint of 200 °C, from which it is observed that the temperature reached the target value at 1000 s, and the temperature fluctuation was approximately ±10 mK, resulting in a negligible variation in the alkali-metal atomic density.

Moreover, the residual magnetic field, which induces transverse spin relaxation, was also measured in situ inside the near zero magnetic shielding on the basis of atomic spin resonance response [28]. The experimental results were depicted in Fig.4(b), and these data were fitted using a linear function. The residual flux density of about 0.134 nT/mA can be derived from the slope of the linear curve. Owing to the optimal structural design and process parameters, these achievements were comparable or even superior to previous results [29, 30].

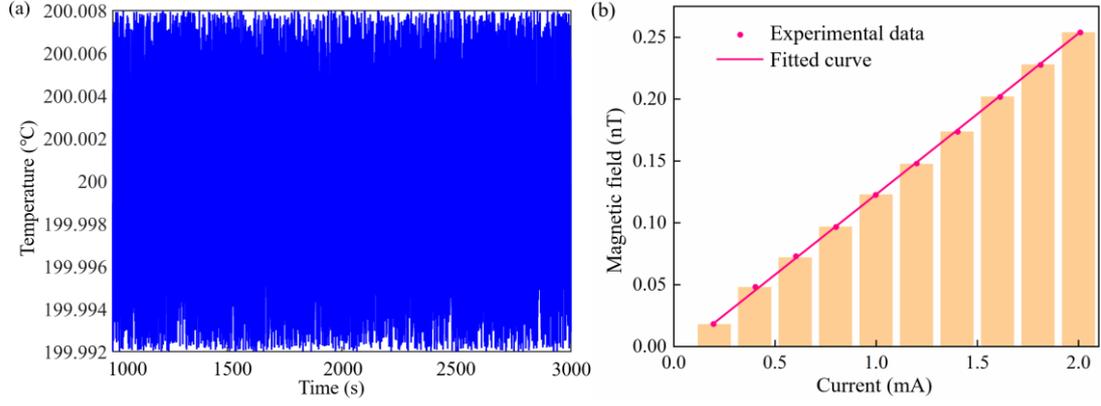

Fig. 4. Performance test of the non-magnetic temperature control chip. (a) Temperature fluctuation at 200 °C; (b) Magnetic flux density generated by the heating chip.

3.1. Optical absorption spectrum

We then measured the absorption spectrum of the alkali-metal atoms, providing compelling evidence for the presence of vapor atoms and buffer gas inside the vapor cell. The shape of the absorption spectrum profile depends on three physical mechanisms: pressure broadening due to collisions between alkali-metal atoms and buffer gas, Doppler broadening due to atomic thermal motion, and natural broadening. For a buffer gas-filled vapor cell, the collision-induced homogeneous broadening dominates, manifesting itself as a simple Lorentzian line-shape. The absorption spectrum of the Rb D1 transition is presented in Fig. 5. The total light attenuation can be quantified using the optical depth (OD) metric:

$$\text{OD} = \ln \frac{P_{in}}{P_{out}} = n r_e c f l \sum_{F,F'} A_{F,F'} \frac{\Gamma/2}{(v-v_0)^2 + (\Gamma/2)^2} \quad (1)$$

where $P_{in}$ is the input optical power, $P_{out}$ is the output optical power, $n$ is the atomic density, $r_e$ is the classical radius of the electron, $c$ is the speed of light, $l$ is the internal length of the cell (i.e. the thickness of the silicon wafer), $f$ is the oscillator strength, $A_{F,F'}$ is the normalized transition strength of the hyperfine transitions, $\Gamma$ is the full width at half maximum of the optical linewidth, and $(v-v_0)$ is the detuning of the laser frequency.

The experimental parameters, including the pressure-shift resonance frequency $v_0$, pressure-broadening linewidth $\Gamma$ and atomic density $n$, can be derived by fitting the experimental data to the model formulated in Eq. (1). The fitted results reveal that the linewidth value $\Gamma$ was approximately 16.38 GHz, corresponding to a buffer gas density of 0.92 amg. This represents a loss of only approximately 8% of the buffer gas initially charged into the bonding cavity during the second anodic bonding. This slight difference is attributable to the non-uniform temperature distribution due to the heating applied during the anodic bonding sealing process.

To further validate the sealing performance of the fabricated integrated vapor cell, we conducted aging tests on 10 wafer-level cells for 30 consecutive days. The tests were performed at a high temperature of 200 °C and a vacuum of approximately 1 Pa, with the optical absorption spectrum measured every five days under the same experimental conditions. The results, presented in the inset of Fig. 5, clearly reveal no noticeable degradation in the pressure-broadened linewidth Γ, indicating good hermeticity.

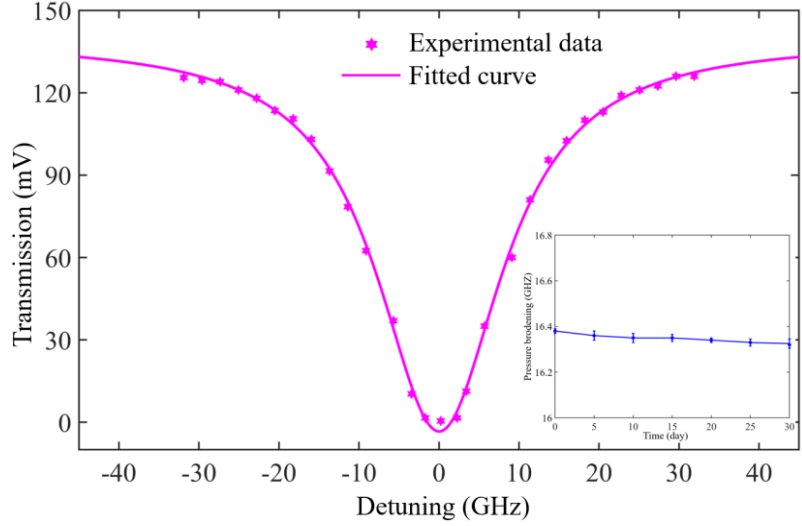

Fig. 5. Optical absorption spectrum of CMOS-integrated AVCs. The power for incident linearly polarized light is adjusted to be as low as possible. The pentagrams indicate the experimental data, whereas the solid blue line indicates the fit using the model formulation in Eq. (1). Inset: variation in the spectral linewidth under 30 consecutive days of extreme testing.

3.2. Saturation absorption spectrum

The SAS is a Doppler-free type of atomic spectrum commonly used for laser frequency locking. Typically, alkali-metal atoms are encapsulated in glass-blown vacuum cells in SAS experiments [31-33]. Likewise, we also clearly observed the SAS of the Rb atom D1 line in the developed microfabricated MEMS cell benefiting from the ultra-thick silicon core (i.e. optical access), as shown in Fig. 6. This experimental phenomenon can be interpreted in the following three aspects:

(1) Given the double-pass configuration with opposite optical propagation paths, only those atoms with zero-velocity components along the light propagation direction can simultaneously interact with both beams when the laser frequency is swept to the resonance frequency of the atomic hyperfine energy level. Consequently, a peak appears on the envelope of the Doppler spectrum, with a frequency corresponding to the hyperfine transition frequency.

(2) The cross-SAS mechanism occurs because a group of atoms with a specific speed in the optical path experiences a resonance transition between different hyperfine energy levels when the laser frequency is exactly at the average value of the resonance frequencies of a certain two pairs of hyperfine energy levels, even though the two beams perceived by the atoms do not share the same frequency.

(3) None of the atoms with any velocity component in the optical propagation path can interact with two beams simultaneously when the laser frequency is at a threshold that is neither at the resonance frequency nor at a value intermediate to the resonance frequencies of the two pairs of hyperfine energy levels.

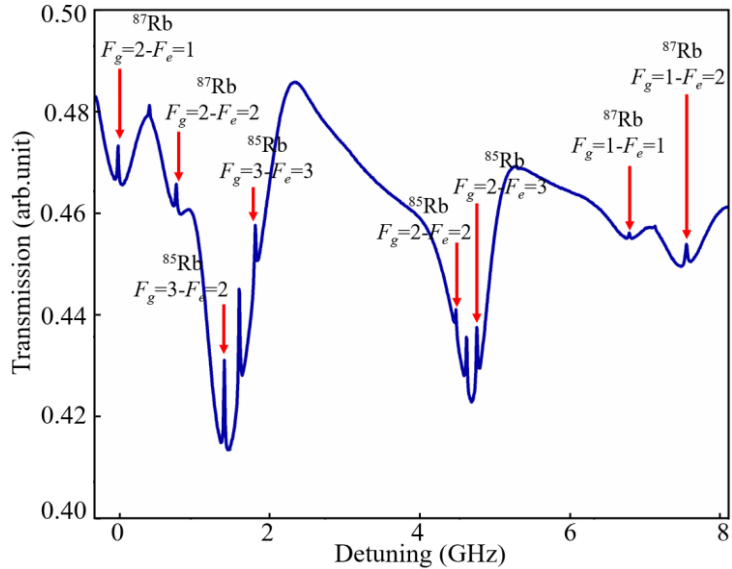

Fig. 6. SAS of the Rb atoms D1 transition line at room temperature. The SAS and cross-SAS are distinctly visible. $F_g$ and $F_e$ are the hyperfine energy levels of the ground and excited states, respectively.

It should be also pointed out that in the SAS experiment, we used a vacuum integrated cell with a vacuum level below $10^{-6}$ mbar, i.e., a molecular pump for vacuuming the bonding chamber was employed during the second anodic bonding.

3.3. Spin noise spectrum

The SNS involves a physical phenomenon where the intrinsic fluctuations of a spin system are in thermal equilibrium. In recent decades, the SNS has emerged as a formidable tool for characterizing the physical properties of unperturbed spin systems derived from the noise power spectrum. However, nearly all SNS experiments have been conducted using conventional glass-blown vapor cells since the spectral signals are very weak [34, 35]. In this study, the spin noise spectrum of unperturbed Rb atoms was visualized using the developed ultra-thick MEMS cells, based on the optical configuration shown in Fig. 3(b). The MEMS cell used for the SNS was filled with naturally abundant Rb atoms ($^{85}$Rb: 72.15%, $^{87}$Rb: 27.85%) and 100 Torr of $N_2$ as the buffer gas. The laser frequency was detuned to -100 GHz with respect to the centerline of the Rb atomic hyperfine transition, allowing for the nondestructive readout of the spin noise. And a transverse magnetic field $B$ = 10 μT perpendicular to the propagation direction of the laser beam was applied to drive the atomic spin precession. Fig.7 illustrates a typical atomic SNS, with the two peaks corresponding to the horizontal coordinates of the Larmor precession frequencies of the natural Rb atoms in the external magnetic field $B$, which are approximately 47 kHz and 70 kHz, respectively.

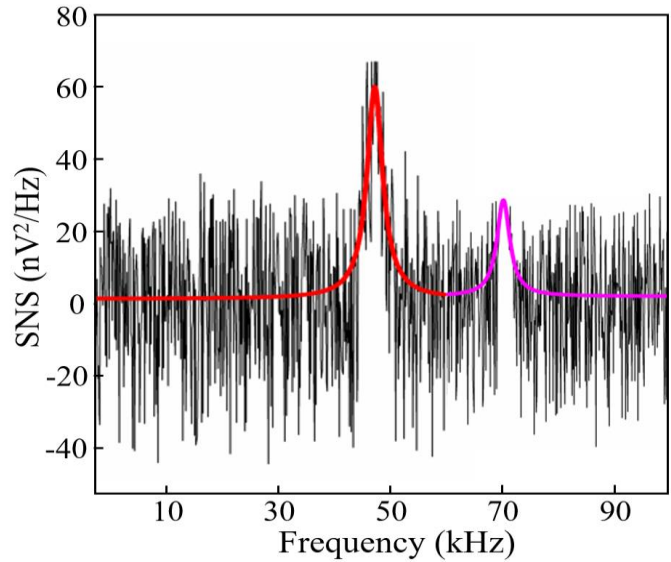

Fig.7. Typical Rb atomic SNS in thermal equilibrium. The black line indicates the experimentally recorded SNS data, the red curve denotes the Lorentzian fitting of the $^{85}$Rb SNS, and the blue curve corresponds to the Lorentzian fitting of the $^{87}$Rb SNS.

The mapping of spin stochastic fluctuations onto the rotation of the polarization plane of linearly polarized light can be interpreted in terms of the circular birefringence effect of Faraday rotation. The magnetic field $B$ splits the Zeeman energy levels in the ground state of the Rb atoms, leading to fluctuations in the absorption coefficients of the atomic ensemble for both the left- and right-handed circularly polarized components.

Therefore, the polarization plane of the linearly polarized light passing through the MEMS atomic vapor cell exhibits a weak rotation angle with respect to the incident light. This rotation is then subjected to a fast Fourier transform (FFT) analysis, yielding the SNS.

3.4. Zero-field-resonance atomic magnetometer

The performance of the CMOS-integrated atomic vapor cell for the single-beam atomic magnetometer was characterized using the experimental setup shown in Fig. 3(b). The heating resistor was used to heat the cell at a driving frequency of 100 kHz, and the temperature sensor measured the temperature in real time. A digital PID controller, programmed for closed-loop temperature control using the LabVIEW software on a PC, maintained temperature stability better than ± 10 mK at 160 °C.

Magnetometers operating under the SERF mechanism yield magnetic resonance based on the ground state Hanle effect, an absorptive measurement that manifests as a change in transmitted power when the atomic ensemble is subjected to a varying transverse magnetic field. Fig. 8 exhibits a representative zero-field magnetic resonance signal as a function of the transverse magnetic field $B_x$, while keeping the other two components, $B_y$ and $B_z$, zeroed. These experimental data can be fitted with Lorentzian or dispersive profiles:

$$P_z^{in-phase}(B_x) = A_0 \left( \frac{\Delta B^2}{(B_x - B_{x0})^2 + \Delta B^2} \right) + C_0 \qquad (2)$$

$$P_z^{out-of-phase}(B_x) = A_1 \left( \frac{B_x - B_{x0}}{(B_x - B_{x0})^2 + \Delta B^2} \right) + C_1 \qquad (3)$$

where $B_{x0}$ is the residual magnetic field in the $x$ direction, $\Delta B$ is the zero-field resonance linewidth denoting the total atomic relaxation rate, and $A$ and $C$ are the system background offset and the resonance signal amplitude, respectively. Following the fitting results, we can derive a magnetic resonance linewidth of 10.5 nT, corresponding to a total relaxation rate of approximately 1800 s$^{-1}$, under 50% electron spin polarization rate. This implies that the optical pumping and total collisional relaxation rates were 900 s$^{-1}$, respectively.

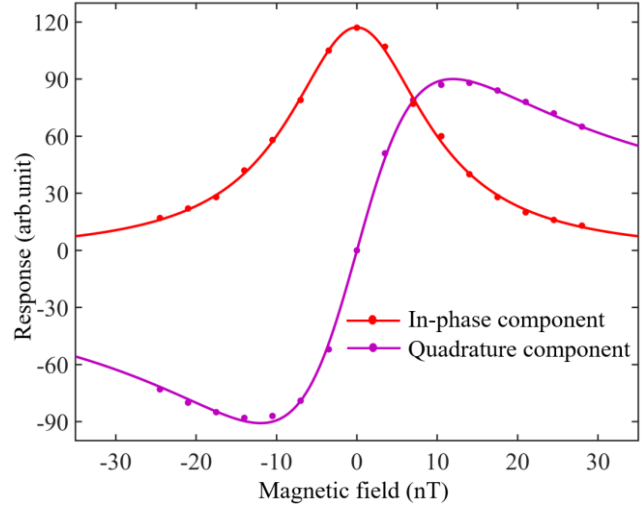

Fig. 8. Zero-field resonance signal as a function of the transverse magnetic field $B_x$ under 50% spin polarization. The red line is the in-phase component fitted by Eq. (2), whereas the purple line is the quadrature component overlaid by Eq. (3).

To operate as a single-beam atomic magnetometer, a sine-modulated magnetic field applied in the $x$ direction converts the zero-field resonance curve into a steeply dispersive profile centered at $B_x = 0$. In our experimental configuration, the frequency and amplitude of the transverse modulated magnetic field were determined to be 890 Hz and 160 nT, respectively, optimized by the scale factor. The transmitted light passing through the ultra-thick vapor cell was demodulated using a lock-in amplifier, yielding the quadrature component of the first-harmonic signal, which was proportional to the transverse magnetic field to be detected. We adopted a straightforward method for magnetic sensitivity calibration. The acquired quadrature component was treated using power spectral density; subsequently, the scale factor converted the output voltage signal into magnetic field information. The magnetic field sensitivity $S_B(f)$ was then calculated as:

$$S_B(f) = \left(\frac{dv}{dB}\right)^{-1} \frac{S_v(f)}{|R(f)|} \approx \frac{\Delta B}{A_{amp}} \frac{S_v(f)}{|R(f)|} \qquad (4)$$

where $dv/dB$ is the slope of the dispersive curve, $S_v(f)$ is the observed amplitude spectral density from the lock-in quadrature component, $R(f)$ is the normalized frequency response, and $A_{amp}$ is the peak-to-peak height of the dispersive curve.

An optimal sensitivity of 12 fT/Hz$^{1/2}$ was achieved in the atomic magnetometer using the aforementioned calibration method. And the electronic noise originating from the photodetector, lock-in amplifier, and data acquisition system was approximately equivalent to 2 fT/Hz$^{1/2}$.

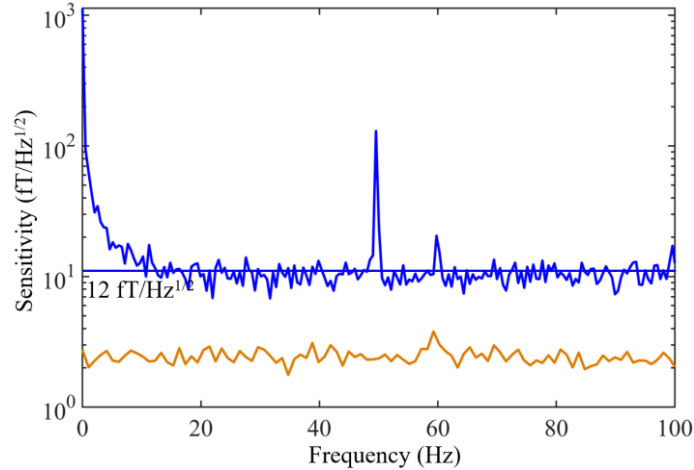

Fig. 9. Noise spectral density of the ultra-sensitive atomic magnetometer. The blue curve indicates the magnetic sensitivity of 12 fT/Hz$^{1/2}$; the brown curve shows the electronic noise recorded during the laser switch-off, including the dark current noise of the photodetectors, lock-in amplifier, and the data acquisition system.

## 4. Discussion

### 4.1 Ultra-long optical path and enhanced performance

The 5-mm optical length, nearly four times longer than that of conventional microfabricated AVCs, directly addresses the critical trade-off between miniaturization and sensitivity in quantum sensing. By utilizing double-sided dry etching and low-temperature anodic bonding, the extended optical path enhances light-atom interaction, enabling the observation of quantum effects such as the SAS and SNS, previously exclusive to macroscopic glass-blown cells. This achievement not only validates the compatibility of MEMS processes with high-performance quantum devices but also bridges the gap between lab-scale prototypes and scalable manufacturing.

### 4.2. Robust hermeticity and functionalized integration

The 30-day aging test under harsh conditions (200 °C, 1 Pa) confirms the exceptional hermeticity of the fabricated AVCs, with only an 8% buffer gas loss attributed to localized thermal gradients during anodic bonding. The wafer-scale integration of non-magnetic heaters and temperature sensors directly onto transparent windows further eliminated the reliance on bulky external components. Such all-in-one functionality is pivotal for applications requiring cost-effective, chip-scale and high-performance situations., such as in bio-magnetic imaging or portable quantum navigation systems.

### 4.3. Novel quantum phenomena and enhanced sensitivity

The demonstration of the SNS in the developed AVCs underscores their capability to resolve weak spin fluctuations in thermal equilibrium, traditionally limited to glass-blown cells. Furthermore, the zero-field atomic magnetometer achieved an ultra-high sensitivity of 12 fT/Hz$^{1/2}$, rivaling that of state-of-the-art quantum magnetic sensors. This performance stems from the synergy between the ultra-long optical access and optimized experimental parameters. These results substantiate the potential of MEMS-based AVCs to replace conventional glass cells in high-precision magnetometry.

**4.4. Challenges and future directions**

While the buffer gas loss and residual magnetic flux density were minimized through process and structural optimizations, further improvements could focus on advanced room-temperature bonding to mitigate gas leakage. Additionally, integrating anti-relaxation coatings (e.g., paraffin and OTS) into the room-temperature bonding process may prolong spin coherence lifetimes, thus enhancing magnetic field sensitivity. Scaling the fabrication process to larger wafers while maintaining yield and consistency remains a practical hurdle, necessitating refinements in etching uniformity and defect control. Finally, exploring hybrid architectures, such as AVCs coupled with waveguide chips or photonic circuits, could unlock new functionalities for multi-modal quantum sensing.

**5. Conclusion**

In summary, we successfully developed and characterized the CMOS-integrated, wafer-level AVCs with a 5-mm silicon core optical length, enabling ultra-high sensitivity quantum magnetometry at 12 fT/Hz$^{1/2}$. The MEMS-based AVCs exhibited robust hermeticity under extreme aging tests and integrated non-magnetic heaters and sensors at the wafer scale, thereby enabling all-in-one functionality. Key quantum phenomena, such as the saturation absorption spectrum and spin noise spectrum, were observed, bridging the performance gap between microfabricated and glass-blown cells. These advancements highlight the potential for scalable, chip-scale quantum sensors in bio-magnetic imaging and other applications of extremely weak magnetic measurements.

# Disclosures

The authors declare no conflicts of interest.

## Data availability statement

The experimental data underlying the results presented in this paper are not publicly available at this time but may be obtained from the corresponding authors upon reasonable request.

## Acknowledgments

This work was supported in part by the National Key Research & Development (R&D) Program of China (Grant No. 2024YFB3212500), National Natural Science Foundation of China (Grant No. 52435010), and the Shaanxi Provincial Science and Technology Development Program (Grant Nos. 2023-LL-QY-35, 2024RS-CXTD-19).